\documentstyle[prl,aps,epsfig,twocolumn, amstex]{revtex}

\begin{document}

 \wideabs{\title{Continuous Variable Quantum Cryptography \\
  - beating the 3 dB loss limit}
 \author{Ch. Silberhorn $^{1}$, T. C. Ralph $^{2}$, N. L\"utkenhaus $^{1}$,
 and G. Leuchs $^{1}$}
 \address{1. Zentrum f\"ur Moderne Optik, Universit\"at Erlangen -
 N\"urnberg, 91058 Erlangen, Germany\\
 2. Centre for Quantum Computer Technology,  University of
 Queensland, QLD 4072, Australia
 \\}

 \date{\today}
 \maketitle

 \begin{abstract}

 We demonstrate that secure quantum key distribution systems based
 on continuous variables implementations can operate beyond the
 apparent 3 dB loss limit that is implied by the beam splitting
 attack \cite{gro01}. The loss limit was established for  standard
 minimum uncertainty states such as coherent states. We show that
 by an appropriate postselection mechanism we can enter a region
 where Eve's knowledge on Alice's key falls behind the information shared between
 Alice and Bob even in the presence of substantial losses.

 \end{abstract}}

 \narrowtext

 \vspace{10mm}

 The distribution of random  keys for cryptographic purposes can be
 made secure by using the fundamental properties of quantum systems
 such that any interception of the key information can be detected.
 This was first discussed for discrete systems \cite{wie83} and
 experimental demonstrations have been carried out using  optical
 sources, which produce low photon number states \cite{butt98}.
 More recently schemes based on continuous quantum variables have
 been proposed \cite{ral99,gottesman01a,02silberhorn,cer01,gro01},
 where the scheme
 by Gottesman and Preskill \cite{gottesman01a} has been proven to be
 information-theoretically secure.  Apart from
 being of fundamental interest these schemes offer certain
 practical advantages. However, they all share one major
 disadvantage: currently it is thought that the use of continuous
 variable techniques does not allow quantum key distribution (QKD)
 beyond 50 \% loss \cite{gro01}. This severely limits the
 applicability of such schemes.

 The argument leading to this limit is based on an optimal cloning
 approach for optical signals that corresponds to a beam splitting
 attack on the signals \cite{gro01}. At the loss limit an
 eavesdropper Eve can replace the lossy channel by a perfect one
 with an adapted beam splitter to mimic the losses. She can then
 generate a cloned signal with a fidelity which depends on the
 beam splitter transmission. In order to extract a secure key out
 of the material with the usual privacy amplification tools
 \cite{bennett95a} and a free choice of the required error
 correction technique, including the efficient two-way schemes
 \cite{brassard94a}, however, the mutual information $I_{AB}$
 between Alice and Bob  has to exceed the information that either
 of them shares with Eve: $I_{AB}
 > \max\{I_{AE},I_{EB}\}$. This condition arises as follows: In order
 to perform privacy amplification \cite{bennett95a} one needs to be able
 to estimate Eve's information on the data
 shared by Alice and Bob after error correction. Two-way error correction
 provides additional information to Eve in
 two forms: redundant information to enable error correction, that are according
 to Shannon's theorem at least
 $1-I_{AB}$ bits, and information about the positions of bits where Alice's and Bob's
 data initially differ.  In
 the worst case, the two-way error correction scheme leaks the complete information
 about these error positions to
 Eve, so that Eve's information about Alice's and Bob's key now stands on the same
 footing and satisfies
 $I'_{(AB)E}\geq \max\{I_{AE},I_{EB}\}$. In the protocol presented below, we actually find
 equality, as explained
 later. Taking this into account, the usual mechanism of  one-way communication schemes applies,
 and we find for
 individual attacks via \cite{bennett95a,csiszar78a,mau93,cachin97a} the condition
 $I_{AB}> \max\{I_{AE},I_{EB}\}$.\footnote{Following discussions with P. Grangier, it should be
pointed out that specific two-way error correction
 techniques might leak less information about error positions leading to less demanding conditions.} Yet, for
 losses beyond $50\%$, one finds that the condition $I_{AB}
 > I_{AE}$ is violated so that the above standard methods cannot be
 used without advanced quantum technologies such as quantum memories and entanglement purification which are
 presently not available. Note that one may  to restrict the information flow for error correction from Bob to
 Alice or vice versa. In this case the ''maximum'' in the above case may safely replaced by the ''minimum''
 \cite{csiszar78a,gro02}, but
  efficient protocols for one way error correction close to the Shannon limit for typical error rates around $5\%$
  are still missing up to
  now.

 In this paper we propose a novel scheme, which operates beyond
the apparent 3 dB limit that is applied by the beam splitting
 attack. In certain situations it is still possible to create a
 secure key \cite{bennett95a} although $I_{AB}<I_{AE} $ or even
 $I_{AB}<I_{BE} $. For classical correlations the procedure is
 known as advantage distillation \cite{mau93}; upon closer
 investigation this turns out to be a form of postselection and
 requires two-way classical communication. Gottesman and Lo
 \cite{gottesmanquant} used this technique to increase the
 tolerance against errors in the single-photon BB84 protocol.
 Postselection is a standard intrinsic procedure in conventional
 QKD with weak pulses: if no photon is detected by Bob, or when
 Alice and Bob did not measure in the same basis, the
 corresponding time slot is ignored and hence does not contribute
 to the raw data. Without this postselection the condition $I_{AB}
 \geq \max\{I_{AE},I_{EB}\}$ could never be reached  for any QKD
 protocol for losses beyond $3$ dB, because then Eve has access to
 better signals than Bob. However, postselection allows
 unconditionally secure key exchange in presence of large losses,
 limited basically only by the photo-detection process \cite{NL}.
 The situation becomes more subtle for continuous variable
 schemes, since then always a non-vacuum signal reaches Bob and
 correlations appear between the data measured by Bob and that of
 an potential eavesdroppers via Alice's state preparation. Thus
 the postselection has to be made more conscious, and here we show
 how to do this. The selection of favorable data for Alice and Bob
 has been previously addressed in the context of implementing the
 BB84 protocol with weak coherent pulses in the presence of a
 strong phase reference pulse \cite{00hir}. Our results
 demonstrate that continuous variables and weak coherent pulse
 schemes  are closely linked in the basic principles.

 We consider the following scheme, which is similar to those
 proposed by Cerf et al. \cite{cer01} and Grosshans et al.
 \cite{gro01}. Alice sends an ensemble of coherent states to Bob
 with a Gaussian distribution of complex amplitudes centered on the
 vacuum. Bob measures either of two conjugate quadratures, say for
 example  the in- and out-of phase quadratures $X$ and $Y$, using
 homodyne detection. The measurement results $x$  are then
 given as  eigenvalues of quadrature operator
 \mbox{$\hat{x}_{\lambda}= \frac{1}{2}(\hat{a} e^{- i \lambda}
 +\hat{a}^{\dag} e^{i \lambda})$} with $\lambda=0$, or
 $\frac{\pi}{2}$. Bob will effectively see a Gaussian distribution
 of real amplitude coherent states when he looks at the quadrature
 $X$ and a Gaussian distribution of imaginary amplitude coherent
 states when he looks at the
  quadrature $Y$. Bob reveals which quadrature he measured
 in each time interval and estimates whether Alice prepared a
 coherent state with a positive or negative displacement in the
 corresponding quadrature. Alice and Bob can now interpret positive
 displacements as logical $"0"$ and negative ones as logical~$"1"$.

 For our analysis of the security of this scheme we extend this
 protocol and specify the used states  by additional steps. After
 Bob's publication of his choice of the quadrature, Alice will
 interpret the state she sent either as a member of the set $\{ |-
 \alpha \ e^{- i \theta }\rangle,  |\alpha \ e^{i \theta }\rangle
 \}$, if Bob detected the quadrature $X$, or $\{ |- i \alpha \ e^{-
 i \theta }\rangle,  |i \alpha \ e^{i \theta }\rangle \}$ ($\alpha
 \in \mathbb{R}$), if Bob measured the quadrature $Y$. She now
 publishes the values of $\alpha$ and $\theta$. In each case, from
 Bob's and Eve's perspective, this narrows down the number of
 possible signals to two, for example  $|\alpha \ e^{ i \theta
 }\rangle$ or $|-\alpha \ e^{-i \theta }\rangle$. Thus Alice and
 Bob can build up a secret key as before when now the encoding
 reads more specifically: $|\alpha \ e^{i \theta }\rangle \to "0
 "$, $|-\alpha \ e^{-i \theta }\rangle \to "1 "$ for amplitude
 measurements and $|i \alpha \ e^{i \theta }\rangle \to "0 " $,
 $|-i \alpha \ e^{-i \theta }\rangle \to "1 " $ for phase
 measurements. Other choices of signal sets are possible, for
 example sets with point symmetry, but the choice above turns out
 to be favorable.

 To investigate the secrecy of the key the distribution of Bob's
 data conditioned on the choice of Alice  can be accessed using
 classical communication. For this purpose  Alice and Bob open up
 complete signal descriptions and measurement results for some
 randomly  chosen transmission events. Thus the statistics of Bob's
 detected results should mirror Alice's coherent state preparation
 with expected Gaussian distributions centered according to the
 complex amplitude displacements.

 Eve's first strategy is thus passive intervention via the
 beamsplitter attack \cite{gro01}. Eve's intervention is
 indistinguishable  from  loss. In the
  typical loss model Alice's state is
 transformed as
 \begin{equation}
 |\alpha \ e^{i \theta }\rangle_{B} |0 \rangle_{E}
 \to  |\sqrt{\eta} \ \alpha \ e^{i \theta }\rangle_{B}
 |\sqrt{1-\eta} \ \alpha \ e^{i \theta }\rangle_{E}
 \label{transform}
 \end{equation}
 for arbitrary $\alpha$ and $\theta$, where $\eta$ is the
 transmission efficiency between Alice and Bob. Alice and Bob are
 none the wiser but Eve ends up with all the lost signal. It was
 shown \cite{gro01}, that provided the loss is less than 50\% it is
 still possible for Alice and Bob to distill a secure key when
 faced with such an attack. We will now show that in fact 50 \%
 loss is not an ultimate limit for secure QKD.

 We wish to find a way by which Alice and Bob can postselect a
 subset of the data for which they have a high mutual information,
 but for which Eve and Alice do not. Alice and Bob can base their
 selection procedure on the parameter $\alpha$ and $\theta$
 characterizing the state preparation and Bob's measurement results
 $x$ or $y$. The overall mutual information of Alice and Bob can
 then be subdivided into different \emph{effective information
 channels} characterized by the parameters $(\alpha, \theta, x)$,
  such that
 \begin{equation}I_{AB}^{tot} =
  \int_{\alpha,\theta, x} dx \  d\alpha \ d\theta \  p(\alpha,\theta, x)
 \cdot I_{AB}(\alpha,\theta, x).
 \end{equation}
 Similarly the overall information  Alice shares with Eve can be
 composed from all single events.

  Note,
 that  the separable nature  of the state  of Eq.~(\ref{transform})
 ensures that there is no correlation between Bob's and Eve's
 quantum uncertainties. Thus  Eve's mutual information with Alice
 does not depend on Bob's detected outcome $x$.  Furthermore, Eve
 shares with Bob always less information than with Alice, $I_{BE}<
 I_{AE}$, and it is sufficient to consider Alice and Eve's
 information only.  Altogether, this allows us to evaluate the
 knowledge of the different parties separately for all effective
 information channels and we can restrict our analysis to find
 suitable parameters $(\alpha, \theta,x)$ with
 $I_{AB}(\alpha,\theta,x)>I_{AE}(\alpha,\theta)$. Since the beam
 splitting attack and protocol itself are symmetrical in respect to
 the considered conjugate quadratures, it is also sufficient to
 investigate  only the case of a quadrature measurement $X$ by Bob.

 To identify the good effective channels, we calculate the mutual
 information shared by Alice and Eve. Alice sends a priori pure
 states. However knowing nothing about Alice's state preparation
 Eve would have to distinguish between two allowed mixed states
 characterized by positive or negative displacements in the
 respective quadrature. So far no general expression for the
 accessible information is known for non-orthogonal mixed states.
 For this reason  we provide Eve with the additional information
 about $\alpha$ and $\theta$. As a trade-off Eve has to
 distinguish for each effective channel between  two
 non-orthogonal \emph{pure} states, in the  case of  $X$
 quadrature measurements between $|\alpha \ e^{i \theta } \rangle$
 and $|-\alpha \ e^{-i \theta } \rangle$. In this situation the
 maximum accessible information is known. It is given as a
 function of the overlap $f$ of the
  respective two states \cite{fuchs} in the form
 \begin{eqnarray}
 \lefteqn{I_{AE}(\alpha, \theta)=}\nonumber\\
 &=&\frac{1}{2}(1+\sqrt{1-f^2(\alpha, \theta)})
 \log(1+\sqrt{1-f^2(\alpha, \theta)}) + \nonumber\\
  & & +\frac{1}{2}(1-\sqrt{1-f^2(\alpha, \theta)})
  \log(1-\sqrt{1-f^2(\alpha, \theta)}). \label{IAE}
 \end{eqnarray}

 For an effective channel with parameters $\alpha$ and $\theta$ the
 overlap can be calculated as
 \begin{equation} \label{overlap}
  f(\alpha,\theta)
  =  \langle -\alpha \sqrt{1-\eta}  e^{- i \theta } | \alpha \
 \sqrt{1-\eta} e^{ i \theta } \rangle\nonumber
  = e^{-2 \cdot (1-\eta) \cdot E^2 },
 \label{prob}
 \end{equation}
 where we defined as $E= \alpha \cos(\theta)$. The protocol
 ensures that the overlap and thus the mutual information of Eve
  depends only on the effective amplitude $E=\alpha \cos(\theta)$.
 As we will see later, the parameters $\alpha$ and $\theta$ enter
  the mutual information of Alice and Bob always in the same
  combination. This allows us to consider in the following the parameters
 $(E,x)$ only.

 At this point we note that the states between which Eve has to distinguish
 and their a-priori probabilities do not
 change if we give Eve the additional information whether Alice's and Bob's
 decoded bit differs for a given signal.
 This kind of information is leaked in two-way error correction (see above).
 It is for this reason that we can
 assume equality in the bound $I'_{(AB)E}\geq \max\{I_{AE},I_{EB}\}$ of Eve's
  information $I'_{(AB)E}$ on Alice's
 and Bob's key given the knowledge of all error positions.

 Next, we calculate the mutual information of Alice and Bob.
 According to the protocol Bob performs quadrature measurements and
 decodes the bit value as the sign of the detected displacement.
 Depending on the signal states $|\pm \alpha \ e^{\pm i \theta }
 \rangle$, his outcomes $x$ are distributed corresponding to one of
 the probability distributions
 \begin{equation}
P( x   | \ |\pm \alpha \ e^{\pm i \theta } \rangle  )
 =|\langle x_0 |\pm \alpha \ e^{\pm i \theta } \rangle|^2
  = \sqrt{\frac{2}{\pi}} e^{-2(x \mp \sqrt{\eta} \
 E ))^2 } \label{prob2}
 \end{equation}
 with  $|x_0 \rangle $ as the eigenstate of the quadrature
 operators with   $\lambda = 0$. This decoding leads to an error
 rate
 \begin{equation} \label{error}
 p_e= \left\{ { \frac{P(x  | |-\alpha \ e^{- i \theta } \rangle
 )}{P(   | |\alpha \ e^{ i \theta } \rangle )+P(x | |-\alpha \ e^{
 -i \theta } \rangle )}  \hspace{0.5cm} \mbox{for }  x>0 \atop
 \frac{P(x  | |\alpha \ e^{- i \theta } \rangle )}{P(x | |\alpha \
 e^{ i \theta } \rangle )+P(x | |-\alpha \ e^{ -i \theta } \rangle
 )}\hspace{0.5cm} \mbox{for }  x<0 } \right.
\end{equation}

 \begin{figure}
  \begin{center}
  \epsfxsize=0.8\columnwidth
  \epsfbox{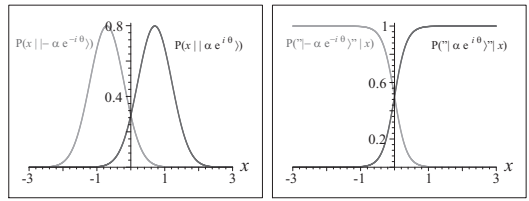}
  \end{center}\caption{Illustration of the error rate of
  Bob's decoding scheme, here for a loss of  $50 \%$ and an
  effective amplitude $E=\alpha\cos(\theta)=1$.
  Left: probability distributions of Bob for prepared bits "0" and "1";
  Right: probability
 that Alice has encoded "0" or "1", if Bob obtained a result $x$.}
 \end{figure}

 To illustrate Bob's decoding we consider in Fig.~1 the case of $50
 \% $ loss and an effective amplitude of $E=1$. The left graph
 depicts the two possible distributions for Bob's results $x$
 corresponding to Alice's states $|\alpha \ e^{ i \theta }\rangle$
 and $|-\alpha \ e^{-i \theta }\rangle$. For the intersection
 point at $x=0$, Bob's measurement delivers no information, but for
 all other values of $x$, the distribution corresponding to Alice's
 encoded bit value is more probable. In the right graphs we
 plotted the probability that Alice has encoded a "0" or "1", if
 Bob actually obtained the value $x$. With the arranged bit
 assignment the smaller value of these graphs gives Bob's error
 probability  corresponding to  Eq.~(\ref{error}). Alice's and
 Bob's mutual information can now be calculated separately for all
 effective information channels with ($x, E$)  by the Shannon
 formula.
\begin{eqnarray}
 I_{AB}(x, E)=1+p_e\log_{2}p_e+ (1-p_e)\log_{2}(1-p_e) \; .
    \label{mB}
 \end{eqnarray}

\begin{figure}
  \begin{center}
  \epsfxsize=0.6\columnwidth
 \epsfbox{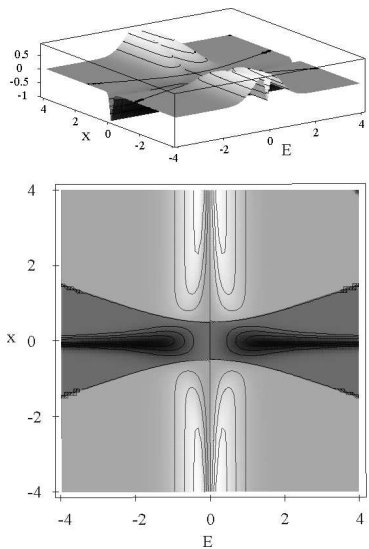}
  \end{center}
 \caption{Comparison of the mutual information between Alice and
 Bob and the information they share with Eve for different state
 preparations with effective amplitude $E=\cos(\alpha)$ and
 measured outcomes $x$ of Bob. Positive areas, colored bright,
 indicate  effective information channels that contribute to secure
 key exchange since they satisfy $I_{AB}(x,E)> I_{AE}(x,E)$.}
 \end{figure}

 We are now in a position to identify those  effective information
 channels with $I_{AB}(x,E)> I_{AE}(x,E)$, which allow to extract a
 secret key. Fig.~2 displays the differences of the respective
 information plotted for the events $(E,x)$ again in the case of
 $50 \%$ loss. Positive valued areas, indicating region of possible
 secure key exchange, are colored bright, negative ones in dark.
 Thus our investigation allows us to model an ideal postselection
 procedure, where all events $(x,E)$ with $I_{AB} >
 \max\{I_{AE},I_{EB}\}$   actually contribute to the key.

 The comparison of Fig.~2 between the mutual information of Alice
 and Bob and the information they share with Eve  dplays an
 insight, that was  first recognized in \cite{00hir}.   Alice and
 Bob can actually utilize statistical measurement results with
 large $x$ to increase their security, but Eve, on the other hand,
 cannot improve her error rate for Bob's selected data. This is,
 because her state is uncorrelated to Bob's results $x$.
 Furthermore there exist for each transmission $\eta$ an optimum
 effective displacements $E$ , such that the mutual information
 between Alice and Bob is maximized.

 We can evaluate the key rates $R_k$, that can be achieved in the
 presented postselection process as
 \begin{eqnarray}
 R_k = R_r \times \int_{S} dx \ dE \ p(x,E) \cdot [
 I_{AB}(x,E)-I_{AE}(x,E)],
 \end{eqnarray}
 where $R_r$ is the raw data rate, and $S$ denotes
 the subset of selected effective channels. For the presented
 protocol the probability $p(x,E)$ that the effective channel is
 used is composed of Gaussian distribution of the effective
 amplitude $E$ and the distribution of $x$ conditioned on $E$.
 With a distribution width
  $d$ for the effective amplitudes we find
 \begin{eqnarray}
  \lefteqn{p(x,E)=}\ \nonumber\\&=& \sqrt{\frac{2}{d \pi}} e^{-2\cdot E^2/d}
  \frac{1}{2}(P(x | |\alpha \ e^{ i \theta } \rangle )+P(x  |
 |-\alpha \ e^{ -i \theta } \rangle ))
 \end{eqnarray}
 with $P(x | |\pm \alpha \ e^{ i \theta } \rangle )$ given in
 Eq.~(\ref{prob2}).

 First numerical calculations, where we limited our integration
 over the data set within $\pm 4$, indicate that in the presence of
 $50 \%$ loss and for an optimized parameter of $d=2.1$, bit rates
 up to $R_k=R_r \times 0.0667$ are achievable. For a loss rate of
 $75 \%$, we have found a key rate of $R_k=R_r \times 0.0073$.
 These calculations show that the high repetition rates of
 continuous variable technology, which is expected to be in the
 GHz region, can actually lead to secure key rates that are  well
 above currently implemented schemes.

 We have shown that continuous variable QKD using coherent states
 in the presence of losses above 50\% can still be implemented
 securely against an eavesdropper using an individual beam
 splitter attack. The postselection process solely relies on
 classical data processing and thus does  not require sophisticated
 quantum resources other than coherent states. An absolute proof
 of security would require the analysis of a more general attack
 by Eve. However, it is likely that the beam splitting attack is
 the optimal attack even in this protocol utilizing postselection.
 The existence of optimum effective displacements for the mutual
 information between Alice and Bob opens the possibility construct
 more elaborated protocols  with modified probability
 distributions for Alice's state preparation   to achieve higher
 the bit rates. However, in all such protocols one has to be
 cautious to ensure that the beam splitting attack remains the best
 eavesdropping strategy.

 Our result pushes continuous variable QKD closer to practical applications.
 In addition we have also shown that the type of
 postselection protocol considered here can also be employed for
 schemes using squeezed light, such as in \cite{02silberhorn}

 This work was supported by the Deutsche Forschungsgemeinschaft and
 by the EU grant under QIPC, project IST-1999-13071 (QUICOV); TCR
 acknowledges the support of the Australian Research Council. The
 authors would also like to thank N. Korolkova for  fruitfull
 discussions and H. Coldenstrodt-Ronge for help in the preparation
 of the manuscript.

 \end{document}